# Evaluation of Machine Learning Algorithms for Intrusion Detection System

Mohammad Almseidin[*], Maen Alzubi[*], Szilveszter Kovacs[*] and Mouhammd Alkasassbeh[§]

[*] Department of Information Technology, University of Miskolc, H-3515 Miskolc, Hungary

[*] Email: alsaudi@iit.uni-miskolc.hu, alzubi@iit.uni-miskolc.hu, szkovacs@iit.uni-miskolc.hu

[§] Information Technology Department
Mutah University, Amman, Jordan
E-mail:mouhammd.alkasassbeh@mutah.edu.jo

*Abstract*—Intrusion detection system (IDS) is one of the implemented solutions against harmful attacks. Furthermore, attackers always keep changing their tools and techniques. However, implementing an accepted IDS system is also a challenging task. In this paper, several experiments have been performed and evaluated to assess various machine learning classifiers based on KDD intrusion dataset. It succeeded to compute several performance metrics in order to evaluate the selected classifiers. The focus was on false negative and false positive performance metrics in order to enhance the detection rate of the intrusion detection system. The implemented experiments demonstrated that the decision table classifier achieved the lowest value of false negative while the random forest classifier has achieved the highest average accuracy rate.

## I. INTRODUCTION

With the rapid development of information technology in the past two decades. Computer networks are widely used by industry, business and various fields of the human life. Therefore, building reliable networks is a very important task for IT administrators. On the other hand, the rapid development of information technology produced several challenges to build reliable networks which is a very difficult task. There are many types of attacks threatening the availability, integrity and confidentiality of computer networks. The Denial of service attack (DOS) considered as one of the most common harmful attacks.

The aim of DOS attacks is to temporarily deny several services of the end users. In general, it usually consumes network resources and overloads the system with undesired requests. For this reason DOS acts as a large umbrella for all types of attacks which aim to consume computer and network resources. [1] In 2000 Yahoo was the first victim of a DOS attack and in the same date also DOS recorded its first ever attack publicly. At the present time, web services and social websites are target of DOS attacks [2]. From another perspective, the remote to local (R2L) attacks are another umbrella for all types of attacks which are designed to have local right permissions because the availability of some network resources is only unique for the local users e.g. file server. There several are types of R2L attacks e.g. SPY and PHF, these types of attacks aim to prepare illegal access to the network resources [3].

As it relates to illegal access to the network and computer resources, User to Root (U2R) attacks aim to switch the attacker access permission from normal user to the root user who has full access rights to the computers and network resources [4]. The main challenge is that attackers are always keeping novelty in their tools and techniques in exploitingany kind of vulnerabilities. Hence, it is very difficult to detect all types of attacks based on single fixed solutions. For that intrusion detection system (IDS) became an essential part of network security. It is implemented to monitor network traffic in order to generate alerts when any attacks appear. IDS can be implemented to monitor network traffic of a specific device (host intrusion detection system) or to monitor all network traffics (network intrusion detection system) which is the common type used.

In general, there are two types of IDS (anomaly base or misuse base). Anomaly intrusion detection system implemented to detect attacks based on recorded normal behavior. Therefore, it compares the current real time traffics with previous recorded normal real time traffics, this type of intrusion detection system is widely used because it has the ability to detect the new type of intrusions. But from another perspective, it registers the largest values of false positive alarm, which means there is a large number of normal packets considered as attacks packets. However, misuse intrusion detection system is implemented to detect attacks based on repository of attacks signatures. It has no false alarm but at the same time, the new type of attack (new signature) can succeed to pass-through it.

Regarding the literature [5] attacks detection considered as classification problem because the target is to clarify whether the packet either normal or attack packet. Therefore, the model of accepted intrusion detection system can be implemented based on significant machine learning algorithms. In this paper, the following implemented the machine learning algorithms have been Implemented (J48, Random Forest, Random Tree, Decision Table, MLP, Naive Bayes, and Bayes Network) to evaluate and accurate the model of intrusion detection system based on a bench market dataset Knowledge Discovery in Databases (KDD) which includes the following types of attacks (DOS, R2L, U2R, and PROBE).

The rest of the paper is organized as follows: section (II) illustrates the related work relevant to using KDD dataset for implementing machine learning algorithms and shows how KDD dataset is very useful. The details steps of reprocess KDD dataset presented in section (III) followed by brief overview of selected machine learning classifiers that is used in the experiments in section (IV). The first phase of building training models experiments is presented in section (V). In section (VI) it describes the evaluation metrics used to evaluate the performance of the selected classifiers used; also it discusses the experiments and the achieved results. Finally, Section (VII) concludes the paper.

## II. RELEVANT WORKS TO THE KDD DATASET

This section presents the related works relevant to using KDD dataset for implementing machine learning algorithms. It also provides a brief overview of the different machine learning algorithms and shows how the KDD dataset is very useful for evaluating and testing various types of machine learning algorithms. The classifier selection model proposed by [2] the authors made a deep survey of intrusion detection system and KDD dataset. They extracted 49596 instances of KDD dataset to implement several machine learning algorithms e.g. Naive Bayes and multi-layer perceptron. Authors succeeded to propose two models for detecting intrusions types of KDD dataset. In [6] the authors implemented support vector machine (SVM) algorithm against network intrusions using MATLAB software. They used KDD dataset as a bench market dataset for intrusions detections. They mentioned that SVM algorithm needs long training time and as a result of that the usability of SVM is limited.

According to another study [7], the authors imported the KDD dataset and implemented the preprocess phases e.g. normalization of the attributes range to [-1, 1] and converting symbolic attributes. Neural network feed forward was implemented in two experiments. The authors have concluded that neural network is not suitable enough for R2L and U2R attacks but on the other hand, it was recorded acceptable accuracy rate for DOS and PROBE attacks. As it relates to implement neural network against KDD intrusions, the effort of [8] the authors succeeded to implement the following four algorithms: Fuzzy ARTMAP, Radial-based Function, Back propagation (BP) and Perceptron-back propagation-hybrid (PBH). The four algorithms evaluated and tested for intrusions detection the BP and PBH algorithms recorded highest accuracy rate.

From another perspective, some of the researchers focus on attributes selection algorithms in order to reduce the cost of computation time. In [9] the authors are focused on selecting the most significant attributes to design IDS that have a high accuracy rate with low computation time. 10% of KDD was used for training and testing. They implemented detection system based on extended classifier system and neural network to reduce false positive alarm as much as possible. On the other hand in [10] the information gain algorithm was implemented as one of effective attributes selection. They implemented multivariate method as linear machine method to detect the denial of service intrusions.

In addition, the genetic algorithm was implemented to enhance detection of different types of intrusions. Meanwhile in [3] a methodology to detect different types of intrusions within the KDD is proposed. The proposed methodology aims to derive the maximum detection rate for intrusion types, at the same time achieved the minimum false positive rate. The GA algorithm used to generate a number of effective rules to detect intrusions. They succeeded to record 97% as accuracy rate based on this methodology. In some cases, if the single isolated machine learning algorithm used to handle all types of intrusions it would be derived by an unaccepted detection rate. In [11] the author used Naive Bayes algorithm to detect all intrusions types of KDD. He illustrated that the detection rate was not acceptable based on single machine learning algorithm.

There are some researchers focusing on specific type of attack such as [12] the authors proposed a system to collect new distributed denial of service dataset which includes the following types of attacks ( http flood , smurf , siddos and udp flood) after the new DDOS dataset proposed. They implemented various machine learning algorithms to detect DOS intrusions, MLP algorithm recorded highest accuracy rate of 98.36%.

All of the previous research works had a respected contributions and at the same time present how the KDD dataset provides the requested environment for testing and evaluation various machine learning algorithms. Also the previous works present that the single isolated machine learning algorithm would not propose the accepted detection rate. In this work, the following machine learning classifiers (J48, Random Forest, Random Tree, Decision Table, MLP, Naive Bayes, and Bayes
Network) were implemented, tested and evaluated based on KDD dataset. The interest is in the most important performance parameters e.g. false negative and false positive to evaluate the selected classifiers. As a result of the implemented experiments the focus will be on selecting the effectiveness ofthe machine learning classifier which achieved the accepted accuracy rate with the minimum false negative value.

## III. KDD DATASET PREPROCESSING AND ANALYSIS

KDD dataset gave a good understanding of several intrusion behaviors, in the same time it is widely used in several areas for testing and evaluation intrusion detection algorithms. The first publicized of KDD dataset was 1999 by MIT Lincoln labs at University of California [13]. It includes 4898431 instances with 41 attributes. In this work KDD dataset was imported to the SQL server 2008 to implement various statistical measurements values e.g. distribution of instances records, attacks types and occurrence ratios. Fig.1. presents the main steps of the KDD dataset import.

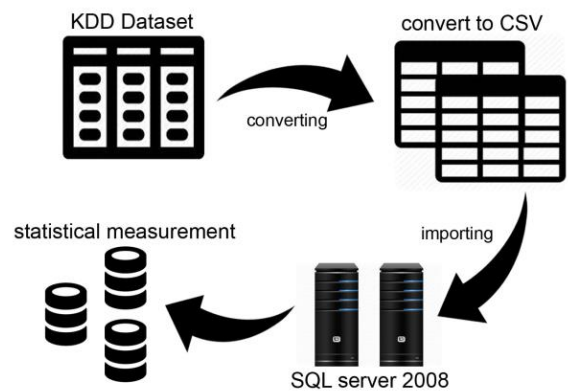

Fig. 1. KDD Dataset Imported Procedure.

Statistical measurements provide a deep understanding of this dataset in order to extract impartial experiments. Table I illustrates the distribution of attacks types within KDD dataset. It can be concluded that there are 21 type of attacks categorized into four groups with different number of instances and occurrences in the KDD dataset. The DOS attacks present 79% of KDD dataset while normal packets present 19% and other attacks types recorded 2% of existing. Based on these values the KDD dataset appears as an unbalanced dataset but at the same time it includes the largest number (41) of packet attributes.

TABLE I. DISTRIBUTION OF ATTACKS WITHIN KDD DATASET.

| Categories of Attack | Attack name | Number of instances |
|---|---|---|
| DOS | SMURF | 2807886 |
| | NEPTUNE | 1072017 |
| | Back | 2203 |
| | POD | 264 |
| | Teardrop | 979 |
| U2R | Buffer overflow | 30 |
| | Load Module | 9 |
| | PERL | 3 |
| | Rootkit | 10 |
| R2L | FTP Write | 8 |
| | Guess Passwd | 53 |
| | IMAP | 12 |
| | MulitHop | 7 |
| | PHF | 4 |
| | SPY | 2 |
| | Warez client | 1020 |
| | Warez Master | 20 |
| PROBE | IPSWEEP | 12481 |
| | NMAP | 2316 |
| | PORTSWEEP | 10413 |
| | SATAN | 15892 |
| normal | | 972781 |

These attributes categorized as a basic information which is collected using any connection implemented based on TCP/IP [4]. Table II illustrates the fundamental attributes information for any connection implemented based on TCP /IP connection environment. The main contribution of this dataset is the introduction of 32 expert suggested attributes which help to understand the behavior of different types of attacks, In other word, the most significant attributes to detect DOS, R2L, U2R and PROBE are included.

TABLE II. THE BASIC ATTRIBUTES OF TCP /IP CONNECTION.

| Attributes | Type |
|---|---|
| Total duration of connections in second | continuous |
| Total number of bytes from sender to receiver. | continuous |
| Total number of bytes from receiver to sender | continuous |
| Total number of wrong fragments | continuous |
| Total number of urgent packets | continuous |
| Protocol type | discrete |
| Type of service | discrete |
| The status of the connection (normal or error) | discrete |
| Label (1) if the connection established from to the same host. Otherwise label (0) | discrete |

IV. BRIEF OVERVIEW OF MACHINE LEARNING CLASSIFIERS

This section provides a brief overview of the different machine learning algorithms and shows the needs to implement machine learning algorithms in various areas such as intrusions detection. The consequence of continued development of technologies makes the need of machine learning algorithms to become more necessary to analyzing and extracting knowledge from a large number of produced datasets. In general; machine learning algorithms can be categorized as supervised algorithms and unsupervised algorithms [14]. Supervised algorithms learns for predicting the object class from pre-labeled (classified) objects. However, the unsupervised algorithm finds the natural grouping of objects given as unlabeled data. In this work, the interest is with the following supervised learning algorithms; because the imported KDD dataset includes the predefined classes.

Multi-layer Perceptron (MLP) Classifier: is one of the most common functions classifiers that prove its effectiveness to deal with several application areas e.g. time series, classification and regression problems. [15] The testing phase can be implemented within short period of time. On the other hand, the training phase is typically implemented in a long period of time. MLP algorithm can be implemented with various transfer functions e.g. Sigmoid, Linear and Hyperbolic. The number of outputs or expected classes and number of hidden layers are important design considerations of the MLP algorithm implementations. At the beginning, every node within the neural network had its randomly weight and bias values, the large weight values present the most effective attributes within a dataset, and on the contrary, the small weight values present the lowest effective attributes within a dataset.

Random Tree Classifier: is one of tree classifiers using this classifying the number of trees should be fixed before implementing. Each individual tree represent a single decision tree. Each individual tree has randomly selected attributes from dataset. Therefore the random tree classifier could be considered as a finite group of decision trees. The procedure of predicting the entire dataset is to migrate several decision trees outputs and choose the winner expected class based on total numbers of votes [16].

Random Forest Classifier: is one of the classification trees algorithms, the main goal of this algorithm is to enhance trees classifiers based on the concept of the forest. Random forest classifiers produced by the referred research [17], had an accepted accuracy rate and can be implemented to handle noise values of dataset. There is no re-modification process during the classification step. To implement this algorithm the number of trees within the forest should be figured because each individual tree within a forest predicts the expected output and after that the voting technique used to select the expected output that have the largest votes number [17]

J48 Classifier: this classifier is designed to improve the implementation of the C.4.5 algorithm which is implemented by Ross Quilan [18] in 1993. The expected output based on this classifier is in the form of decision binary trees but with more stability between computation time and accuracy [19]. Regarding to decision tree structure the leaf node had a decision of expected output.

Naive Bayes Classifier: this classifier refers to the group of probabilistic classifiers. It implements Bayes theorem for classification problems. The first step of Naive Bayes classifier is to determine the total number of classes (outputs) and calculate the conditional probability for each dataset classes. After that the conditional probability would be calculated for each attribute. The standard formula of Naive Bayes can be found in the referred research [8]. Furthermore, it has the ability to work with discreet and continuous attributes also on the contrary of MLP classifier Naive Bayes can be implemented within a short period of time [11]. meanwhile Naive Bayes can be represented as Bayesian network (BN) or Belief network. BN supports presenting independent conditional probability based on understanding framework. In general BN is acyclic graph between expected class (output) and a number of attributes [20].

Decision Table Classifier: the main idea of this classifier is to build a lookup table, it helps to identify the predicted class of output. There are several search algorithms e.g. breadth first search, genetic algorithm and cross validation can be implemented to generate the efficiency of the decision table [21]. The lookup table includes a set of conditions and the expected actions refer to the predefined conditions. To put it in another way; the consequence of decision table classifier is set of significant rules help to predict the new incoming inputs [22]. The lookup table of the decision table can be used on other area e.g. it can be used to present the significant rules for the fuzzy system when the system is complex and there is a lack of expert knowledge base.

## V. TRAINING MODELS DATASET EXPERIMENTS

Regarding to the KDD dataset there are 21 type of attacks categorized into four groups (DOS, R2L, U2R, and PROBE) with different number of instances and occurrence in dataset. After the KDD dataset imported to SQL server 2008. 148753 instances of records have been extracted and presented in Table III as training data. Based on a deep analysis of KDD dataset the distribution occurrence of different types of attacks was saved. In other words 79% of extracted dataset present DOS attacks and 19% for normal traffic while 2% for other types of intrusions (U2R, R2U and PROBE).

TABLE III. TRAINING MODEL DATASET.

| Categories of Attack | Attack name | Number of instances |
|---|---|---|
| DOS | SMURF | 85983 |
|  | NEPTUNE | 32827 |
|  | Back | 70 |
|  | POD | 10 |
|  | Teardrop | 30 |
| U2R | Buffer overflow | 10 |
|  | Load Module | 2 |
|  | PERL | 1 |
|  | Rootkit | 5 |
| R2L | FTP Write | 2 |
|  | Guess Passwd | 10 |
|  | IMAP | 4 |
|  | MulitHop | 2 |
|  | PHF | 1 |
|  | SPY | 1 |
|  | Warez client | 31 |
|  | Warez Master | 7 |
| PROBE | IPSWEEP | 382 |
|  | NMAP | 70 |
|  | PORTSWEEP | 318 |
|  | SATAN | 487 |
| normal |  | 28500 |

In this paper, the experiments were performed on Ubuntu 13.10 platform, Intel R, Core(TM) i5-4210U CPU @ 1.70GHz (4CPUs), 6 GB RAM. Waikato Environment for Knowledge Analysis (WEKA) is a machine learning tool written in JAVA [23]. It is an open source tool and available for free. The numerical classification examples appearing in this paper are provided by the WEKA toolbox. The Most common machine learning classifiers are used in this experiments (J48, Random forest, Random Tree, Decision Table, Multilayer Perceptron (MLP), Naive Bayes and Bayes Network). Based on 148753 instances of records it was successful to create the training models for all the selected machine learning classifiers. All the studied models are prepared and compared for a comprehensive study of machine learning classifiers efficiency.

## VI. MACHINE LEARNING CLASSIFIERS EXPERIMENTS, RESULTS AND DISCUSSION

After the creation of the training models, the next step is the testing phase process implementation. In order to implement a fair testing phase fully randomized 60000 have been extracted. The extracted testing data includes all 21 types of attacks within KDD dataset. There are several evaluations metrics can be used in a classification algorithm. In this paper, the confusion matrixes were generated for each machine learning classifiers. It includes significant information about existing and predicted output classes. Furthermore, the following performance metrics are computed [12]:

- True Positive (TP): this value represents the correct classification attack packets as attacks.

- True Negative (TN): this value represents the correct classification normal packets as normal.

- False Negative (FN): this value illustrates that an incorrectly classification process occurs. Where the attack packet classified as normal packet, a large value of FN presents a serious problem for confidentiality and availability of network resources because the attackers succeed to pass through intrusion detection system.

- False Positive (FP): this value represents incorrect classification decision where the normal packet classified as attack, the increasing of FP value increases the computation time but; on the other hand, it is considered as less than harmful of FN value increasing.

- Precision: is one of the primary performance indicators. It presents the total number of records that are correctly classified as attack divided by a total number of records classified as attack. The precision can be calculated according to the following equation:

$$P = \frac{TP}{(TP+FP)} \quad (1)$$

In addition, the number of both the correctly and the incorrectly classified instances are recorded with respect to the time taken for proposed training model. During the testing phase, the following parameters were applied for the machine learning classifiers. J48 tree classifier was tested with confidence factor = 0.25; numFolds = 3; seed = 1; unpruned = False, collapse tree = true and sub tree rising =true. Random forest classifier also tested with number of trees =100 and seed =1. Random tree classifier was tested with min variance = 0.001 and seed = 1. A decision table classifier was tested based on the Best
First Search (BFS) and cross value = 1. Furthermore, the MLP classifier was tested with the following parameters: search learning rate=0.3, momentum =0.2, validation threshold=20.

Table IV presents the TP rate and the Precision values of the selected classifiers in the experiments. It can be concluded that the random forest classifier achieved highest TP rate of 93.1% while the random tree classifier achieved the lowest TP rate of 90.6%. In other words, random tree classifier reached the lowest value of attacks classification process. Form another perspective, the decision table classifier reached the lowest precision value of 94.4% and that indicates the decision table classifier suffers of an increasein false positive value. Therefore, there is a large number of normal packets classified as attack packets.

TABLE IV. TRUE POSITIVE RATE AND PRECISION RATIOS.

| Machine Learning Classifiers | TP Rate | Precision |
|---|---|---|
| J48 | 0.931 | 0.989 |
| Random forest | 0.938 | 0.991 |
| Random tree | 0.906 | 0.992 |
| Decision table | 0.924 | 0.944 |
| MLP | 0.919 | 0.978 |
| Naive Bayes | 0.912 | 0.988 |
| Bayes Network | 0.907 | 0.992 |

In general, TP rate and precision values are important performance parameters for a common intrusion detection system, but from another perspective the most serious performance parameters are FP rate and FN rate. The research works of intrusion detection system aim to decrease both of these parameters as much as possible; specifically, the FN parameters. According to Fig.2. which illustrates the FP and FN performance parameters, it can be concluded that the random tree classifier achieved the highest FN rate of 0.093. Hence there is a large number of attacks classified as normal packet. On the contrary, the decision table classifier is achieved the lowest FN rate of 0.002. In the same time, the decision table classifier reached the highest FP rate of 0.073 and that means there is a large number of normal packet classified as attack packets.

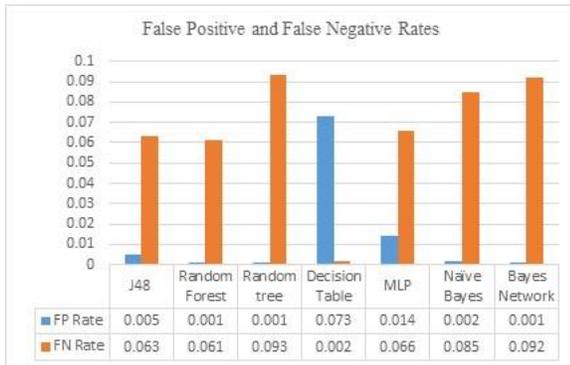

Fig. 2.    False Positive Rate and False Negative Rate.

Table V presents the Root Mean Square Error (RMSE) and area under the Receiver Operating Characteristic (ROC). RMSE presents the difference between the actual and the desired outputs based on the confusion matrix. The model which has lower RMSE is a more efficient than a model having a larger RMSE. Meanwhile ROC value calculated based on true positive and false positive. The large value of ROC indicates the ability of a model to detect intrusions while the lower value present the weakness of a model.

TABLE V. ROOT MEAN SQUARE AND AREA UNDER THE RECEIVER OPERATING CHARACTERISTIC.

| Machine Learning Classifiers | ROC Area | Root Mean Squared Error |
|---|---|---|
| J48 | 0.969 | 0.0763 |
| Random forest | 0.996 | 0.0682 |
| Random tree | 0.953 | 0.0763 |
| Decision table | 0.984 | 0.0903 |
| MLP | 0.990 | 0.0813 |
| Naive Bayes | 0.969 | 0.0872 |
| Bayes Network | 0.997 | 0.0870 |

Regarding Table VI that Bayes network classifier recorded the highest value 0.999 based on ROC value while random tree classifier presented as lowest value 0.953. Furthermore, random forest classifier had the lowest value 0.0682 based on RMSE indicator while the decision table presented as highest value 0.0903. Through the testing and classification of 60000 instance of records from the KDD dataset. The total number of incorrectly classified records for each selected classifiers are presented in the Table VI. The average accuracy rate is calculated by the following formula:

$$AverageAccuracyRate = \frac{TP+TN}{TP+FN+FP+TN} \quad (2)$$

TABLE VI. AVERAGE ACCURACY RATE.

| Machine Learning Classifiers | Correctly classified Instances | incorrectly classified Instances | Accuracy Rate |
|---|---|---|---|
| J48 | 55865 | 4135 | 93.10% |
| Random Forest | 56265 | 3735 | 93.77% |
| Random tree | 54345 | 5655 | 90.57% |
| Decision table | 55464 | 4536 | 92.44% |
| MLP | 55141 | 4859 | 91.90% |
| Naive Bayes | 54741 | 5259 | 91.23% |
| Bayes Network | 54439 | 5561 | 90.73% |

Another issue could be the time required for building the classifier training models. Based on the experiments random tree classifier built training model in the fastest time, while MLP classifier built its model during 176 minute; which is the longest time. The results of the numerical examples can be concluded in the following points:

- The Random forest achieved the highest accuracy rate 93.77 with smallest RMSE value and false positive rate.

- The Random tree classifier reached the lowest average accuracy rate 90.73 with smallest ROC value.

- Regarding to the average accuracy rate there is no big difference between MLP classifier and Naive Bayes classifier.

- All machine learning classifiers present acceptable precision rates for detecting normal packets.

- Bayes network classifier recorded the highest value for detecting correctly the normal packet.

- There are no big differences between MLP and J48 classifiers based on FN parameters.

- The decision table classifier did not reached the highest accuracy rate, but it had the lowest FN rate and

  it has a low time demand for building the training model.

- All of the selected machine learning classifiers except MLP built their training models in accepted period of times.

- It can be concluded that the group of rules classifiers (the decision table) can present an acceptable

accuracy rate with the lowest FN rate, which also increases the confidentiality and the availability of the network resources.

## VII. CONCLUSIONS

In this paper, several experiments were performed and tested to evaluate the efficiency and the performance of the following machine learning classifiers: J48, Random Forest, Random Tree, Decision Table, MLP, Naive Bayes, and Bayes Network. All the tests were based on the KDD intrusion detection dataset. The rate of the different type of the attacks in the KDD dataset are approximately 79% of DOS attacks, 19% of normal packets and 2% of other types of attacks (R2l, U2R and PROBE). In the experiments 148753 instances of records have been extracted as training data to build the training models for the selected machine learning classifiers.

The testing phase is implemented based on 60000 random instances of records. Several performance metrics are computed (accuracy rate, precision, false negative, false positive, true negative and true positive).

The experiments have demonstrated that there is no single machine learning algorithm which can handle efficiently all the types of attacks. The decision table (rules base classifiers) achieved the lowest false negative value of (0.002), but it was far from the highest accuracy rate detection. On the other hand, Bayes network classifier had the highest value for correctly detecting the normal packets. Random forest classifier registered the highest accuracy rate 93.77%, with the smallest RMSE value and false positive rate. It seems that the random forest classifier presents acceptable performance parameters except the false negative parameter. In contrast, all of the selected machine learning classifiers, except the MLP, were able to built their training models in an acceptable period of time. Furthermore, to save the availability and the confidentiality of the network resources, the true positive and the average accuracy rates alone are not sufficient to detect the intrusion. False negative and false positive rates are also needed to be taken into consideration.

## VIII. ACKNOWLEDGEMENT

The described study was carried out as part of the EFOP3.6.1-16-00011 Younger and Renewing University – Innovative Knowledge City – institutional development of the University of Miskolc aiming at intelligent specialization project implemented in the framework of the Szechenyi 2020 program. The realization of this project is supported by the European Union, co-financed by the European Social Fund.